\documentstyle[12pt]{article}
\makeatletter                    
\@addtoreset{equation}{section}  
\makeatother                     

\begin{document}

\title{Semiclassical Gravity and Mesoscopic Physics
\thanks{
Invited talk  at the International Symposium
on Quantum Classical Correspondence, Drexel University, Philadelphia,
Sept. 8-11, 1994.
To appear in the Proceedings edited by D. H. Feng and B. L. Hu
(International Press, Boston, 1996)}}
\author{B. L. Hu\\
{\small Department of Physics, University of Maryland,
College Park, MD 20742, USA
\thanks{permanent address. e-mail:hu@umdhep.umd.edu}}\\
{\small Institute for Advanced Study, Princeton, New Jersey 08540, USA}\\
{\small  Department of Physics, Hong Kong University of Science and
Technology,} \\
{\small Clear Water Bay, Kowloon, Hong Kong}}
\date{\small {\it (UMDPP 96-50, IASSNS-HEP-95/xxx, October, 1995)}}
\maketitle

\begin{abstract}

Developments in theoretical cosmology in the recent decades show a
close connection with particle physics, quantum gravity and
unified theories. Answers or hints to many fundamental questions
in cosmology like the homogeneity and isotropy of the Universe,
the sources of structure formation and entropy generation,
and the initial state of the Universe can be traced back to the activities
of quantum fields and the dynamics of spacetime from the Grand
Unification time to the Planck time at $10^{-43} sec$.
A closer depiction of this primordial state of the Universe
requires at least a semiclassical
theory of gravity and the consideration of non-equilibrium statistical
processes involving quantum fields. This critical state is intermediate between
the well-known classical epoch successfully described by Einstein's Theory of
General Relativity and the completely unknown realm of quantum gravity.
Many issues special to this stage such as the transition from quantum to
classical spacetime via decoherence, cross-over behavior at the Planck scale,
tunneling and particle creation,  or growth of density contrast from
vacuum fluctuations 
share some basic concerns of mesoscopic physics for condensed matter,
atoms or nuclei, in the quantum/ classical and the micro /macro interfaces,
or the discrete / continuum and the stochastic / deterministic transitions.
We point out that underlying these issues are three main factors:
quantum coherence, fluctuations and correlation. We discuss how a
deeper understanding of these aspects of fields and spacetimes
can help one to address some basic problems, such as Planck scale
metric fluctuations, cosmological phase transition and structure formation, and
the black hole entropy, end-state and information paradox.
\end{abstract}

\newpage
\section{Introduction}

This conference covers four rapidly developing areas of physics: the
foundational
aspects of quantum mechanics, quantum gravity and cosmology, mesoscopic physics
and chaos. Though apparently disjoint, they share the common concern of how
to correctly understand the quantum and classical descriptions with regard to
their individual particularities and mutual consistency. In this talk I'd like
to address two aspects: Quantum cosmology \cite{qg,qc} and mesoscopic physics
\cite{meso}. (Another work on using the concept of decoherence to explore
the relation between quantum and classical chaos is reported by
Shiokawa \cite{ShiHu} in this conference.)
Specifically, I want to show how semiclassical gravity can be viewed as
mesoscopic physics.
By semiclassical gravity I mean the theory where gravity is treated classically
by Einstein's Theory of General Relativity, and matter field quantum
mechanically
by quantum field theory. This is represented by the discipline of
quantum field theory in curved spacetime \cite{BirDav},
which provides a good description of
gravity and matter  below the Planck energy $10^{19}$ GeV
(or Planck time $10^{-43}$ sec, Planck length $10^{-33}$ cm) and constitutes
the theoretical framework for understanding quantum processes in the
early universe and black hole physics from the Planck energy to the
Grand Unified Theory (GUT) energy $10^{15}$ GeV).
Mesoscopic physics deals with problems where the
characteristic interaction scales or sample sizes are intermediate
between the microscopic and the macroscopic. For the experts they refer to a
specific set of problems in condensed matter and atomic / optical physics
(see, e.g., \cite{meso}). For the present discussion, I will adopt a
more general definition, with `meso' referring to the interface between
macro and micro on the one hand and the interface between classical and quantum
on the other. \footnote{
Another meaning of mesoscopia can be defined with respect to structures and
interactions.
Instead of dwelling on these individual processes in their specific context,
one can refer to the general category of problems which exist in between
two distinct levels of matter structure or interaction scales, such as between
the molecular and atomic scales, the QED lepton-hadron and nuclear scale,
the nuclear and particle (quark-gluon) scales, the QCD and GUT
(grand unification  theory) scale (assuming deserts in-between),
and of course, from GUT to QG (quantum gravity) scale,
which is depicted by semiclassical gravity.
The distinct levels of interaction are not arbitrarily picked,
they obey theories of a fundamental (QED, QCD) or quasi-fundamental
(atomic, nuclear interaction) nature. The meso scales between them have common
traits. They usually fall in the range where the approximations taken from
either level (e.g., low energy QCD versus perturbative hadron physics)
fail, and new structure depicted by new collective variables and new language
are called for to study its behavior. The new problems encountered in condensed
matter and nuclear/particle physics fall under such a conceptual category.
So do the problems of extending semiclassical gravity towards quantum gravity
or projecting quantum gravity (e.g., superstring theory) onto low
energy particle physics (the standard model).}
These two aspects will often bring in the continuum / discrete and
the deterministic / stochastic factors.
I will show  how issues concerning the micro / macro interface
and the quantum to classical transition arise in quantum cosmology and
semiclassical gravity in a way categorically similar to the new problems
arising
from condensed matter and atomic/optical physics (and, at a higher energy
level,
particle/nuclear physics, at the quark-gluon and nucleon interface).
I will show that many issues are related to the coherence and
correlation properties of quantum systems, and involve
stochastic notions, such as noise, fluctuations, dissipation and
diffusion in the treatment of transport, scattering and propagation processes.
The advantage of making such a comparison between these
two apparently disjoint disciplines is twofold: The  theory of mesoscopic
processes which can be tested in laboratories with the newly developed
nanotechnology can enrich our understanding of the
basic issues common to these disciplines while being extended to the realm of
general relativity and quantum gravity.
The formal techniques developed and applied to problems in quantum field
theory and spacetime geometry
can be adopted to treat condensed matter and atomic/optical systems with
more rigor, accuracy and completeness.
Many conceptual and technical challenges are posed by
mescoscopic processes in both areas.

\section{Problems in Semiclassical Gravity and Mesoscopic Physics where
Quantum and Statistical Mechanics Play a Fundamental Role}

For many years I have maintained the somewhat unconventional view that
cosmology
and astrophysics should be considered as `condensed matter' physics
\cite{HuHK}.
This refers to  both classical and  quantum processes. \footnote{
Although I have worked on quantum processes at the somewhat remote
Planck scale ($10^{-43}$ sec, $10^{-33}$ cm), and I share the enthusiasm
with other advocates on the importance of explaining most common observations
from their quantum origin (the most ambitious of all in this attempt is the
`getting everything from nothing' philosophy -- structure and dynamics
from quantum noise or vacuum fluctuations)
I do think that many of the observed cosmological phenomena arose from
nonlinear processes (hydrodynamic, kinetic or mechanical, including chaos)
at very late times. I don't think a single stroke, no matter how appealing
the theory appears (e.g., superstring theories purported by those as the theory
of everything TOE, or a certain proposal of the boundary condition of the
wave function of the universe) can explain everything we see in the universe
today.
The appreciation and explanation of the hierarchy of structures and
interactions in the natural world require more in-depth analysis on the
cross-over between each `elementary' level and the composite levels
derived therefrom. See, e.g., \cite{HuSpain}}
(For a recent account of a similar viewpoint, see \cite{smolin}.)
At face value, this statement could be taken to mean that these
subjects are applications of general relativity (GR) and
quantum field theory (QFT)
(for early universe and late stage black hole processes),
as condensed matter physics is with  respect to quantum mechanics and
statistical physics. Less trivially, it is how the viewpoint and approach
to important problems in cosmology and astrophysics which I want to make
a distinction from the way traditionally pursued in GR or QFT.
By nature, cosmology and astrophysics
are interdisciplinary subjects. Unlike GR or QFT and their mathematical
derivatives such as superstring theory, conformal field theory or canonical
gravity via old and new variables, which rely on formal constructions
and abstractions, sometimes even to the extent of defying reality,
these disciplines involve many physical processes and many different effects
to paint even an approximate picture.
In essence, their growth has always been strongly influenced or enriched by
other major currents of theoretical and observational physics.

Looking at the development of theoretical cosmology and astrophysics
in the last forty years,
if one wants to name one important branch of physics in each decade
which enhanced its growth, one could say that in the fifties,
it was nuclear physics which laid down the background for processes
later observed in neutron stars, and explained nucleosynthesis
in the early universe (between 1 sec and 3 minutes). In the sixties,
it was relativity (of course, to begin with,  general relativity was
responsible
for the inception of modern cosmology and astrophysics in the 20's and 30's --
the three solar system tests, Hubble expansion,
Friedmann and Schwarzschild solutions were their cornerstones, followed by
detailed studies of gravitational collapse in the 30's and discoveries of
exact solutions in the 50's ). The observation
of microwave background radiation, quasars and pulsars, which ushered in the
golden age of contemporary relativistic cosmology and astrophysics.
In the seventies, it was quantum field theory and semiclassical gravity theory
which drew the attention of the importance of quantum processes in addressing
some of the most important issues in theoretical cosmology,
such as the singularity and the horizon problems
(possible singularity avoidance and horizon removal due to quantum
processes such as trace anomaly and particle creation \cite{cospar,cosbkrn})
and which provided the theoretical basis for the discovery of the celebrated
Hawking effect \cite{Hawking} in black holes.
In the eighties, it was from particle physics
considerations that inflationary universe \cite{infcos} was introduced,
which quickly became the focus of current  cosmological research activities.
With it, the recognition that the early
universe can provide a cheap and convenient laboratory to test out possible
particle physics theories from the energy range of the Standard Model to that
of the Grand Unified Theories provided further impetus to the establishment
of a new field of particle astrophysics, not unlike the efforts in the fifties
by the nuclear physicists in the establishment of nuclear astrophysics.

What about the nineties? In my opinion, statistical mechanics, especially
nonequilibrium statistical mechanics, kinetic theory, stochastic mechanics,
chaos, fractals, and complexity studies,  will provide
new stimulus and show new avenues for the development of both physical
cosmology  and gravitation physics.

In  recent reviews I have enumerated some important quantum processes
in the early universe  which require a nonequilibrium statistical
mechanical treatment.
I discussed processes such as
phase transition, particle creation, and entropy generation
in the Waseda University Conference on Quantum Physics and the Universe
\cite{HuWaseda}.  In the Los Alamos Conference on Fluctuations and Order
\cite{HMLA}, I have expanded this list and
discussed specifically the relevance of noise and fluctuations to the problems
of decoherence in the transition from quantum to semiclassical
gravity,  particle creation, entropy generation, galaxy formation,
and suggested that the `Birth of the Universe'
be viewed in the light of large fluctuation phenomenon.
Let me display that list and add a few more entries \cite{Banff,jc}:\\

TABLE 1\\

\noindent 1. Cosmological particle creation as
parametric amplification of vacuum fluctuations\\
2. Thermal radiance from accelerated observers, moving mirrors, and black holes
   as fluctuation-dissipation phenomena\\
3. Entropy generation from quantum stochastic and kinetic processes\\
4. Phase transitions in the early universe as  noise-induced processes\\
5. Galaxy formation from primordial quantum fluctuations\\
6. Anisotropy dissipation from particle creation as backreaction processes\\
7. Dissipation in quantum cosmology and the issue of the initial state\\
8. Decoherence, backreaction  and the semiclassical limit of quantum gravity\\
9. Inflationary multiverse network, cellula automata, and complexity\\
10. Percolation, nucleation, and spinodal decomposition\\
11. Topology change in spacetime and loss of quantum coherence problems\\
12. Stochastic spacetime, coarsening and continuum limit\\
13. Spacetime foam, bubble and froth: the kinetics of topology and geometry \\
14. Regge-Ponzano quantum gravity, spin-network \\
15. Gravitational entropy, singularity  and time asymmetry\\
16. `Birth' of the universe as a spacetime fluctuation and tunneling
phenomenon\\

\section{Mesoscopic Physics -- Fundamental Issues at the Quantum/ Classical
and Micro/ Macro Interfaces}

To practitioners in condensed matter and atomic/optical physics, mesoscopia
refers
to rather specific problems where, for example, the sample size is comparable
to the probing scale (nanometers), or the interaction time is comparable to the
time of measurement (femtosecond), or that the electron wavefunction correlated
over the sample affects its transport properties, or that the fluctuation
pattern
is reproducible and sample specific. Let me select a partial list of these
processes defined experimentally or phenomenologically:\\

TABLE 2\\

\noindent 1. Large n Rydberg atom:  quantum chaos. Extent of validity of
quasiclassical
approximations, traces or fingerprints of classical chaos in corresponding
quantum systems.\\
2. Aharonov-Bohm effect and Berry phase. Phases of wavefunctions and topology
of
field configurations. A-B oscillations in loops and conductance fluctuations
in wires.\\
3. Quantum Hall effect: quantum correlations, topology effects of electrons in
lower-dimensional systems.\\
4. Localization:  electron transport in stochastic media. Interference between
diffusion paths, coherent backscattering.\\
5. Quantum transport: Landauer scattering, coherent versus dissipative
transport.
Universal fluctuations in conductance.
Sample specific and reproducible fluctuation patterns.\\
6. Quantum optics: dephasing, quantum trajectories, squeezing and
uncertainty.\\
7. Sonoluminescence: dynamical Casimir effect and micro to macro energy
transfer.\\
8. Discrete to continuum transitions: lattice gauge theory and infrared
limit.\\
9. Finite size effect: correction to universality behavior at critical point
 in bulk systems, correlation length affected by finite size.\\
10. Fluctuations and noise: effect of quantum and thermal fluctuations.
Nonequilibrium noise, fluctuation-dissipation, coherent and incoherent
tunnelling,
resonant tunneling in quantum wells.

\subsection{Mesoscopic Physics}

1. Quantum / classical correspondence, Deterministic / stochastic dynamics\\

This issue rests at the foundation of quantum mechanics. Recent work has shown
how nonequilibrium statistical mechanics can provide deeper insight into the
issue of quantum to classical transition \cite{envdec,conhis}.
A particularly interesting
application of the phenomena of decoherence is to study the relation
between quantum and classical chaos \cite{chaos}. The study of quantum chaos
has been focused
on searching for the clues or traces (scars and fingerprints) of the quantum
counterpart of a chaotic classical system.  How a quantum system evolves or
changes over  into a classical chaotic system is not easily addressed.
It is of interest to study how decoherence of a quantum system by interaction
with an environment or by the action of noise can quantitatively depict
the emergence of classical chaotic behavior \cite{ShiHu,ZurPaz}
and thus provide the missing link between these rather disparate domains.\\

\noindent 2. Micro / macro structures and dynamics \\

This is of course the general theme of statistical mechanics, but the
issues take on a special significance when coupled with the quantum / classical
interface. A well known example of macroscopic quantum processes is
quantum tunneling with dissipation \cite{CalLeg83}. Another example is
sonoluminescence. Viewed as dynamical Casimir effect \cite{SchwingerSL},
the (macroscopic) classical collapse of a dielectric media excites the
(microscopic) quantum fluctuations resulting in the emission of light.
We shall see many semiclassical gravity processes are of this nature.\\

\noindent 3. Medium energy scale, intermediate wave zone \\

This is the scale where existing approximation schemes from either end
(high and low energy, near and far zone, short and long wavelength regime)
via WKB, Born-Oppenheimer, or similar methods fail. It is also the range
where neither few-body dynamics or statistical description can cope with.
In the sense defined in footnote 1, this issue is at the heart of mesoscopia.
It is not just a time-honored problem in mathematical physics in terms
of coming up with techniques to deal with this difficult range, but
it bears on deeper theoretical physics issues, such as the choice or emergence
of a suitable set of collective variables for the best description of the
intermediate scale physics, the conditions for the relative closure (decoupling
and renormalizability)
of effective theories and the  cross-over behavior between the low and high
energy theories. \\

\noindent 4. Discrete / continuum limit and extended / localized
state transitions\\

Examples are numerous in this category, such as localization  of
wave scattering in random media (e.g., Sheng in \cite{meso}),
melting or coarsening transition. Formally there is a correspondence
between the Einstein lagrangian describing the evolution of spacetime
(geometrodynamics) and that describing the dynamics of an interface
\cite{randomgeom,Zia}
(zero cosmological constant corresponds to adding surfactants)
Correspondingly, galaxy formation from  gravitational instability can
be viewed as an interface growth problem \cite{growth}.\\

\noindent 5. Finite size versus bulk behavior\\

This started with phase transition studies in the early 70's concerning
the correction in the infrared behavior of the order parameter fields in
systems of finite size compared to bulk systems \cite{Fisher}.
The central issue is how the transition from micro scales
to macro scales can be understood with the running of coupling constants and
interaction parameters from the ultraviolet into the critical regime,
and how the finite size of the
sample changes  the scaling behavior.\\

\noindent 6. Quantum Transport\\

Traditional transport theory applied to mesoscopic structures is based on
near-equilibrium or linear response approximations (e.g., Landauer formula).
New nanodevice operations involve nonlinear, fast-response and far-from-
equilibrium processes. This recognition calls for new developments in the
theory of transport. One serious approach via the Keldysh method using
Wigner functions (e.g., Buot in \cite{meso}) is closely related
to similar formalisms developed for nonequilibrium quantum fields aimed at
problems in the early universe \cite{CH88}.\\

\noindent 7. Correlations and Fluctuations\\

Again examples are many: Sample specific fluctuation patterns,
universal conductance fluctuations, strongly correlated systems \cite{meso}.

\subsection{Problems in Semiclassical Gravity}

Let us now take a look at problems in semiclassical gravity of a similar
nature.\\

1. A necessary task for any proponent theory of quantum gravity (in addition
to showing its intrinsic viability in, say, addressing the issue of time, etc)
is to demonstrate that it has the correct limit of semiclassical gravity,
or explain how the classical spacetime picture  as depicted by the theory of
general relativity emerges. At the heart of this problem is
decoherence, a necessary but not sufficient condition. Quantum decoherence was
studied for minisuperspace models of  quantum cosmology  since the mid-80's
\cite{Halliwell}.
Amongst the Interesting outcomes were the derivation of the semiclassical
Einstein equations, the proof that the no-boundary or tunnelling boundary
conditions of the wavefunction of the universe \cite{HarHaw,Vil}
can lead to an inflationary epoch, with the de Sitter invariant
vacuum prevailing. The assumptions which go into this transition are discussed
in detail in \cite{PazSin}.\\

2. Semiclassical gravity is at the {\it micro / macro interface}. Here
gravity is treated macroscopically by general relativity, while
matter described by quantum field is by nature microscopic.
Many of the semiclassical gravity processes has this property.
Examples are Casimir energy of quantum fields in curved or topologically
nontrivial spacetimes, cosmological particle creation from vacuum fluctuations
of quantum fields. The latter is, in our assessment,
closely related to sonoluminescense viewed as dynamical Casimir effect,
in that the dynamics of the universe (governed by Einstein's equation)
parametrically excites the vacuum fluctuations of the quantum fields into
particle pairs.  Particle creation provides an energy transfer mechanism
from the macroscopic (geometrodynamics) to the microscopic (particles and
fields).
When the backreaction  of quantum field processes is included,
the dynamics of spacetime is driven
by the expectation value of the energy momentum tensor operator of
the quantum matter field. (This is the reverse, i.e., micro to macro, process.)

What is the proper tool or language for depicting the microscopic structure
of spacetime based on our knowledge of the low energy physics?
The hope is that it is describable by suitable generalizations
of the well-proven quantum field theory (e. g., by string field theory)
in a non-perturbative way (not based on the linearized theory
of spin 2 particle).
Conceptually we have taken a different route here:
moving up in energy scale rather than down, i. e.,
the construction or deduction of the
attributes of a microscopic theory of gravity from analyzing the
fluctuations in the critical regime from the macroscopic theory
(Einstein's general relativity and its generalized semiclassical form).
To highlight the conceptual contrast, we can even say that in certain ways
Einstein's general relativity theory is the hydrodynamic limit of a more
fundamental microscopic theory \cite{GRhydro}.
This `bottom-up' rather than `top-down'
approach we have taken has the simple advantage that one starts  from
a more familiar terrain, that of semiclassical gravity and its fluctuations.
Although we don't entertain the hope of deducing completely the high energy
or microscopic structure of spacetime (we recommend a serious look at
superstring theories \cite{GSW}, despite its seeming retreat,
and the knot-theory representation of quantum relativity of
Ashtekar, Rovelli and Smolin \cite{Ash}), we
think the study of critical phenomena at the Planck scale via the
approach we have taken will reveal some special properties of the
sub-structure.
Difference in viewpoints notwithstanding, it is the next logical step
and a worthwhile challenge. \footnote{
After all, this is the approach physicists have undertaken for centuries,
in extracting new physics of a smaller scale from the contradictions of
old physics at a larger scale -- molecular kinetic theory from thermodynamics,
quantum physics from atomic spectroscopy, particle physics from nuclear
processes, etc.}\\

3. The third issue relating to the treatment of {\it intermediate energy range}
is a common but nontrivial one. In the quantum / classical gravity context,
this refers to the regime where one can no longer assume
the wave function of the universe to be in the WKB form \cite{Calzetta}.
New phenomena are expected to arise here.
This problem exists  in all  decoherence considerations of
quantum to classical transition, i. e., whether the WKB wavefunction
has a classical attribute (no in general -- there is no peaking in the phase
space \cite{HabLaf})
or whether the coherent state has the most classical character
(yes in general -- it is the state which has the lowest entropy \cite{ZHP}).
In terms of viewing classical persistent structures as arising from
 coarse-graining the quantum histories, a more difficult related issue is
 what collective variables \cite{HuSpain}
 will be most suited for the description of
 the intermediate scale physics, and how one constructs
them from the accessible low energy phenomenology.

When considered in the context of
decoherence studies, be it the `choice' of an environment (degree of subjective
or objective prerogative) in the environment -induced approach  \cite{envdec}
or realizing the projection operators in the decoherent history approach
\cite{conhis},  if the Planck scale
is not given but left as an emergent parameter, this will take on the
difficult question of what gives rise to the Planck scale as the regime where
classical behavior emerges, and to what extent the scales of consistent
and stable structures depend on coarse-graining the progenatory quantum system.
If general relativity is viewed as hydrodynamics \cite{GRhydro},
this issue takes on
the question of whether above the Planck scale, there can exist intermediate
scales which admit stable and persistent structures, similar to
the many scales which exist in the kinetic theory regime before the system
settles into the long wavelength hydrodynamics regime \cite{Spohn},
or the existence of resonances as excited states of composite particles,
as witnessed by the profusion of `elementary' particles in the 60's and 70's.\\

4. {\it Discrete and continuum spacetime} Lattice gravity \cite{LatGra}
and simplicial geometry \cite{RegCal}
such as depicted by Regge calculus is an important line of inquiry
bridging discrete models of quantum gravity to classical general relativity.\\

5. {\it Finite size effect} in cosmology refers to the effect of spacetime
geometry
and topology on the infrared behavior of quantum fields near a phase
transition.
O'Connor and I took up this study ten years ago \cite{HuOC}. The
effect is, strictly speaking, neither due specifically to topology nor
curvature,
but the presence of a finite size in the background spacetime or in the
fluctuation field operator.
It has an effect on the nature of phase transition (first or second order)
and how it is approached and consummated (dynamics or energetics).
Here, the role of quantum fluctuations in inducing phase transitions
is of basic importance.\\

6. {\it Nonequilibrium quantum fields and nonlinear quantum transport}.
Calzetta
and I \cite{CH88} had developed a kinetic theory for nonequilibrium quantum
fields, using the Schwinger-Keldysh (closed-time-path) formalism, the
nPI effective action method with Winger functions. (See also \cite{neqfN}
and compare with Buot in \cite{meso}).
This method has been applied to an analysis of  reheating by particle
creation \cite{infreh} and first order phase transitions \cite{cosphatra}
in the inflationary universe , and to heavy-ion
collisions in nuclear reactions \cite{heavyion}.
Recent summary of work in these directions can be found
in meetings of thermal field theory and its applications \cite{tft}. \\

7. {\it Quantum fluctuations} in the inflaton field can be viewed as noise
which
seeds galaxy formation \cite{StoGalFor} and entropy generation.
Quantum origin of noise and
fluctuations has been discussed by \cite{cgea,HuBelgium,nfsg}.
In attempting to deal with more exotic ideas such as the `birth' of the
Universe, one encounters  difficulty even in explaining the meaning of
words such as tunnelling from `nothing' . But in terms of mesoscopic physics,
it is possible to visualize it as a large fluctuation problem
in stochastic processes, or a phase transition problem in critical dynamics.
We think it is important to first understand the physics of these
processes in a more familiar setting before trying to formulate it in
the uncharted waters of quantum cosmology. Mesoscopic physics
is such an area where new physics can be learned and ideas tested out.

\section{Common Points -- Quantum Coherence, Fluctuations, Correlations}

Viewed in a more theoretical light, we can decipher
three aspects which underlie most of the processes named above. They are
quantum coherence, fluctuations and correlations.  All mesoscopic processes
involve one or more of these aspects. \\

\noindent 1. Fluctuations and decoherence: quantum / classical
correspondence:\\

\noindent
Decoherence: Fluctuations and noise in the environment is what is responsible
for decoherence in the system, which is a necessary condition for quantum to
classical transition.\\
Classicality: Classical description in terms of definite trajectories in phase
space requires correlations between conjugate variables. Noise and fluctuations
destroy this correlation. The observed classical reality as an emergent
phenomenon from quantum description has intrinsic stochastic behavior
\cite{GelHar2,HuTsukuba}.\\

\noindent 2. Coherence and dissipation \\

This is the counterpart to 1. above, as fluctuation and dissipation are
balanced
by the fluctuation-dissipation relation. (We found that this relation exists
for general, including non-equilibrium, conditions as it originates from the
unitarity of the closed system from which the open system is defined
\cite{nfsg,fdrsc}).
Coherence of electron wavefunction in a mesoscale sample makes it sensitive
to circuit geometry, as manifest in the detection of Aharonov-Bohm effects.
Anderson localization can be understood as due to coherent back-scattering
from random sources.
Coherence in quantum systems is altered by  dissipative effects,
as occurs in macroscopic quantum phenomena \cite{CalLeg83}, e.g.,
competition between coherence and dissipation manifests in
tunneling at finite temperature.\\

\noindent 3. Correlation\\

Strongly correlated system has very different  transport properties
related to scattering or diffusion processes. Long range order established
near the critical point invokes higher order correlations which
determine the nature of phase transitions. Correlation between particles and
a quantum field is a determining factor for many atomic / optical processes. \\

Let me mention some current research directions in semiclassical
cosmology and black hole physics which also involve these aspects in a
fundamental way.\\

A. The above-mentioned issue of
 decoherence and transition from quantum cosmology to semiclassical
gravity has been treated with the theory of quantum open systems and
the method of influence functional. The recent work of  Calzetta, Matacz,
Sinha and me \cite{nfsg,fdrsc,HM3} (see also \cite{CamVer})
show that an Einstein- Langevin equation
is a natural generalization of the semiclassical Einstein equation at the
quantum gravity transition.
The latter is a mean field theory obtained by taking a noise average
of the former. This provides a new platform for one to investigate
fluctuation processes related to particle creation at the Planck scale. These
work also show that the backreaction of particle creation resulting
in the dissipation in the dynamics of the background spacetime can be
viewed as a manifestation of a fluctuation- dissipation relation. One can
use these results to explore deeper issues such as metric fluctuations
in spacetime, both curved and flat, near the Planck scale. Kuo and Ford
\cite{KuoFor} calculated fluctuations in the Casimir energy  and considered
possible experimental observations.  Nicholas Philips \cite{PhiHu} has
been calculating the fluctuations and variance in the energy momentum tensor
of quantum fields in curved spacetimes. I have speculated that the variance
may come out to be largely independent of the local geometry and topology.
(somewhat similar to the universal fluctuations of conductance observed in
mesoscopic processes, which is independent of sample size and shape).
If this were true then it will reveal some simple yet basic properties
of Planck scale physics.\\

B. Structure formation from quantum fluctuations in the early universe\\

This important problem involves the fluctuation and decoherence aspects.
The current folklore is that the classical fluctuations (noise) arise
from the high frequency quantum fluctuations of
a quantized {\it free} field, with no due considerations of the quantum
origin of noise and the decoherence process.
We have raised serious doubts on such a practice \cite{cgea,HuBelgium,nfsg}.
Recently Calzetta and I have shown
by way of an interacting field theory model (a scalar - gravitational field
coupling) \cite{qfsf} how taking the correct procedure can actually produce
an improved result which alleviates the necessity of assuming an unnaturally
small coupling constant. In the traditional and we think erroneous treatment,
the fluctuations part of the quantum field is implicitly viewed as completely
decohered and acts as a stochastic classical source in the equation governing
the density perturbations. We pointed out that
it is the noise term resulting from the decoherence of the mean field
which enters into the Langevin equation. The quantum fluctuations are
instrumental to the decoherence of the mean field but are themselves not
the direct source of classical noise.
There is an interplay of decoherence and fluctuations here,
which contributes to the density contrast proportional to an extra factor of
the coupling constant (thus in a $\lambda \phi^4$ model for the inflaton, the
$\lambda$ can be as high as $10^{-6}$ rather than
assuming  an unnaturally small $10^{-12}$). \footnote{This can perhaps be
compared to the difference between diamagnetism and paramagnetism,
the former requires the field first to polarize the media before interacting
with it, thus the extra factor of interaction involved.}
This subtle but important point is missed out in
the conventional treatment.\\

C. Correlation dynamics, nonlocality of quantum fields and
black hole information paradox\\

While working on a formalism of nonequilibrium quantum field theory
first from the correlation dynamics (Boltzmann-BBGKY) \cite{CH88} and then
from the open system viewpoint (Langevin) \cite{HPZ}, it became increasing
clear to us that even though sophisticated techniques
(e.g., Feynman perturbation) and concepts (e.g., renormalization theory)
in quantum field theory have been developed,
some simple problems in particle field interaction
have not been understood well enough yet. This is because historically
we are more interested in the energetics aspects of the systems, such
as transition amplitudes, scattering cross sections, energy level shifts
due to interactions of particle and fields.
The statistical mechanics aspects of particle and fields
has not been explore to the same extent.
Problems such as the correlation, noise, fluctuations, and dissipative
dynamics of  a system are traditionally dealt with in statistical mechanics,
but rarely for quantum fields and particle-field interactions. Studies
of quantum processes in strong field and curved spacetime
conditions such
as in the early universe and black holes have led us to such inquiries.
(Other areas such as quark-gluon plasma, heavy ion collisions
and high field atomic / optical processes have also seen such a recent demand.)
Questions such as the noise, dissipation and  entropy a particle- field system
-- how they are related to particle creation and quantum fluctuations on the
one hand and the correlation dynamics and kinetics on the other -- requires
a deeper probe into the stochastic properties of quantum fields.
Even from the limited results our investigations have produced,
one can see the rich physics and potential
this nascent subject of nonequilibrium
field theory holds which describes the statistical mechanics of particles
and quantum field systems.

As sample problems, let me mention the  derivation of
Hawking and Unruh radiation from statistical field theory, viewed
as kinematical effects without invoking {\it a priori} geometric notions or
constructs \cite{Ang,HM2}. The work of Raval, Anglin
and I on the statistical properties of particle (detectors and probes)
- field systems \cite{RHA,RH} and that of Raval, Johnson and I on
the backreaction of particle creation
(Unruh effect) on the accelerated detector and moving mirror problems
are precursors to the problems of black hole entropy,
backreaction and information paradox.
On this issue, I have been of the
opinion that the correlations in the quantum field has a strong role
to play in where the information from the black hole can be stored
and transferred. To see this in a clearer light, Calzetta and I have
constructed a platform to describe the  statistical mechanical properties of
interacting quantum fields, i.e., in terms of the
{\it dynamics of the correlation functions}, after the Boltzmann-BBGKY scheme
\cite{Balescu,CH88}. The main theme contained in two recent papers
\cite{dch,cddn}
can be summarized as follows:
Starting from the thesis that the full dynamics of an interacting
quantum field may be described by means of the Dyson- Schwinger equations
governing the infinite hierarchy of Wightman functions which measure the
correlations of the field, we showed how
this hierarchy of equations can be obtained from the variation of the infinite
particle irreducible, or {\it `master' effective action} (MEA).
Truncation of this hierarchy
gives rise to a quantum subdynamics governing a finite number of  correlation
functions (which constitute the `system'), and expression of the higher
order correlation functions (which constitute the `environment') in terms
of the lower-order ones by functional relations (`slaving' or `factorization')
induces {\it dissipation} in the dynamics of the subsystem driven
by the stochastic
fluctuations of the environment, which we call the {\it `correlation  noises'}.
These two aspects are related by the  fluctuation-dissipation relation.
This is the quantum field equivalent of the BBGKY hierarchy in
Boltzmann's theory. Any subsystem involving a finite
number of correlation functions defines an effective theory,
which is, by this reasoning, intrinsically dissipative.
The relation of loop expansion and correlation
order is expounded. We see that ordinary quantum field theory which involves
only the mean field and a two-point function, or any finite-loop effective
action in a perturbative theory are, by nature, effective theories
which possess these properties.
Histories defined by lower-order correlation functions can be decohered by the
noises from the higher order functions and acquire classical stochastic
attributes.  The present scheme invoking the correlation order is a
natural way to describe the quantum to classical transition for a closed
system as it avoids {\it ad hoc} stipulation of the system-environment split.
It is through decoherence that the subsystem variables become classical
and the subdynamics becomes stochastic.\\

The application of such a conceptual scheme (correlation dynamics of
interacting field) for addressing the black
hole information paradox problem is currently under study. A sketch of
the main ideas (including scaling properties of fields at infrared limit
as induced by the black hole)
are described in \cite{erice}, where the reader will also find discussions of
metric fluctuations and Einstein- Langevin equation.
At a simpler level, recently
Anglin et al \cite{AngRecoh} has used a particle - (free) field model to
illustrate recoherence, and drew implications of  their results
on the black hole information paradox.\\

Here we have only presented a sampling of current problems in
semiclassical gravity which share with mesoscopic physics
some common concerns of the basic issues in both fields.
We hope this sketch can induce more thoughts
and discussions and bring out more interesting insights
beneficial to the development of both disciplines.\\

\noindent {\bf Acknowledgement}
Many ideas I discussed or proposed here are based on results obtained
from work done over the past ten years in collaboration with Esteban Calzetta,
Salman Habib, Andrew Matacz, Denjoe O'Connor, Juan Pablo Paz, Apan Raval,
Kazutomu Shiokawa, Sukanya Sinha, Chris Stephens and Yuhong Zhang. I thank
them for sharing the pleasure and excitement of searches and insights.
(Any misrepresentation or misconception is sheerly due to my
own ignorance or misjudgement.)
I have also benefited from discussions with  Prof. Zhao Bin Zu of
Academia Sinica, China, and  Profs. Ping Sheng,  Zhao Qing Zhang, T. K. Ng
and colleagues and students of the theory group at the physics
department of HKUST.
Research is supported in part by the National Science Foundation
under grant PHY94-21849.  Support from
the General Research Board of the Graduate School of the University of
Maryland and the Dyson Visiting Professor Fund at the  Institute
for Advanced Study, Princeton are also gratefully acknowledged.


\end{document}